\documentclass[12pt]{article}

\usepackage[space]{grffile}
\grffilesetup{
encoding,
filenameencoding=cp1251
}

\usepackage{amsfonts}
\usepackage{amssymb}
\usepackage{amscd}
\usepackage{amsmath}

\newcommand{\BEQ}{\begin{equation}}
\newcommand{\EEQ}{\end{equation}}
\def\bea{\begin{eqnarray}}
\def\eea{\end{eqnarray}}
\def\nn{\nonumber}

\newtheorem{Th}{Theorem}
\newtheorem{Lem}{Lemma}

\newtheorem{Def}{Definition}

\newtheorem{Rem}{Remark}

\def\bea{\begin{eqnarray}}
\def\eea{\end{eqnarray}}

\def\bes{\begin{equation*} \begin{split}}
\def\ees{\end{split} \end{equation*}}

\def\C{{\mathbb{ C}}}

\usepackage{graphicx}
\graphicspath{{pics/}}

\begin{document}
~\\
\centerline
{\LARGE \bf Towards integrable structure in 3d Ising model }
\vskip 10mm 
~\\
\centerline
{\large Dmitry V. TALALAEV\footnote{Geometry and Topology Department, Faculty of Mechanics and Mathematics,
Moscow State University, Moscow, 119991 Russia
E-mail: dtalalaev@yandex.ru} (MSU, ITEP)}
\vskip 15mm
\abstract{
We construct a weight matrix for the 3D Ising model satisfying the so-called twisted tetrahedron equation. The result is based on the theory of the n-simplicial complex and the invented recursion procedure on the space of n-simplex solutions in correspondences. The weight matrix reveals some properties intrinsic for the hypercube combinatorics. }

~\\
\tableofcontents
\vskip 2cm
\section{Introduction}
The Ising model \cite{Bax} is an amazing area of interaction between algebraic and geometric methods, topology and exactly solved models in statistical physics. Among others, it describes critical phenomena in magnets and ice-models. We dare to mention the strong relevance of this model with the Hopfield neural network \cite{Hopf}. 

The principal goal of the paper is the integrability of the 3D system, which remains still hypothetical. We develop an algebraic approach to this problem based on the so-called twisted tetrahedron equation, which could be considered as an analog of the famous Zamolodchikov tetrahedron equation \cite{Zam}. We would be happy to reconstruct the ``big'' commutative family appearing in the statistical models produced with the weight matrix satisfying the original Zamolodchikov equation \cite{T1}. 

We would like to indicate two aspects close to the integrability of the 3D Ising model attracted much attention in last decades: the NP-completeness of the 3D Ising model \cite{NP} and the conformal field theory approach to the solution \cite{CFT}. 

The main part of the work involves the combinatorics of the $n$-simplicial complex \cite{KST}. We first establish a recursion procedure on the spaces of solutions for the $n$-simplex equation. Then we propose such a weight matrix in 3D Ising model, which satisfies the twisted tetrahedron equation with spectral parameter. Despite this equation does not provide the same simple integrability property as the original one we aimed this paper to draw attention of the experts community to this phenomenon. Moreover, the nature of this solution lies in the domain of the hypercube combinatorics and is concerned with fundamental problems in coding theory and parallel computing \cite{har}.

\subsection*{Acknowledgements.} 
I would like to thank the IHES institution and the Angers University where the part of this work was done for hospitality and favored atmosphere. I would like to make special thanks to D. Gurevich, A. Odesskii, V. Roubtsov for stimulating questions. The work was partially supported by the RFBR grant 18-01-00461.

\subsection{Vertex model}
The isotropic Ising model is described by the Hamiltonian 
\bea
H(\sigma)=\sum_{d(i,j)=1} \sigma_i \sigma_j\nn
\eea
where $i\in\Lambda$ - a periodic 3d lattice, $\sigma_i$ is the spin variable associated to the $i$-th vertex. $d(i,j)$ is the standard (Manhattan) metric on the cubic lattice.
The partition function is defined as follows:
\bea
\label{part}
Z(t)=\sum_{(\sigma)}\exp(t H(\sigma))
\eea
where the sum is taken over the space of all spin configurations.

First of all we pass to the variables $s_{ij}$ on edges and associate the space $\C^2$ to each edge of $\Lambda.$ Then we consider the dual lattice $\Lambda^*$ those vertices are 3-cubes of $\Lambda$, the 2-faces of $\Lambda$ are edges of $\Lambda^*.$ We associate a vector space $V_f=(\C^2)^{\otimes 4}\simeq \C^{16}$ to each edge of the dual lattice $\Lambda^*.$ We define a weight matrix $W$ which is interpreted as a linear operator on the space $(V_f)^{\otimes 3}\to (V_f)^{\otimes 3}$ those matrix elements are $0$ is the corresponding spin configuration is not admissible and takes the value
\bea
\exp(t(\sigma_1+\sigma_2+\sigma_3))\nn
\eea
for admissible configurations of spins. Here $\{\sigma_i\}$ is a fixed set of 3 edges of the 3-cube of different direction. A configuration is called admissible if the product of spins over each 2-face equals 1 and if the coloring of different 2-faces with common edges are consistent.
\begin{Rem}
Quite obviously this weight matrix provides the product formula for the Ising model partition function 
\bea
Z(t)=\prod_{(\alpha,\beta,\gamma)\in\Lambda^*} W_{\alpha\beta\gamma}.\nn
\eea
This is due to the fact that each edge of the lattice $\Lambda$ enter the fixed set of three edges of different directions for some 3-cube of the lattice $\Lambda.$
\end{Rem}

\section{Recursion on $n$-simplex varieties}

The aim of this section is to compound a natural recursion on the spaces of solutions for the $n$-simplex equation in correspondences. 

\subsection{$n$-simplicial complex}
Let us remind some notations from \cite{KST} and define the $n$-simplex equation. Let $R: X^{(n+1)}\to X^{(n+1)}$ be a map or a correspondence $R\subset X^{(n+1)}\times X^{(n+1)}$.

\begin{Def}\label{def1}
The admissed coloring of $(n-1)$-faces in a cube $I^N$ with respect to the chosen $R$ is a coloring, such that for all $n$-face $f_n\subset I^N$ the colors of outgoing $(n-1)$-faces are linked with the colors of the incoming $n-1$ faces by $R$. We denote the space of colorings as $C_N^{n-1}(X,R).$ 
\end{Def}

It turns out that this space is not trivial iff 
$R$ satisfies the algebraic equation called the $n$-simplex equation. This is defined uniquely on $(n+1)$-faces without higher syzygies. 

To define this equation let us consider the $(n+1)$ - cube $I^{n+1}$ and the oriented graph $G_{n+1}$ those vertices are $n$-faces of $I^{n+1}$ and the edges are defined as $(n-1)$-faces which are outgoing for one $n$-face and outgoing for another. We associate a direction to such an edge in an obvious way. It turns out that $G_{n+1}$ is a disconnected sum of two graphs which are both $n$-simplexes. 

\begin{Th}\label{th1}
The graph $G_{n+1}$ has two connected components, each isomorphic to an $n$-simplex. One of them (let call it "left") contains faces
\begin{equation}
\label{left}
(0***\ldots*),\quad ({}*1**\ldots*),\quad (**0*\ldots*),\quad \ldots \;.
\end{equation}
The faces:
\begin{equation}
\label{right}
(1***\ldots*),\quad ({}*0**\ldots*),\quad (**1*\ldots*),\quad \ldots \;,
\end{equation}
are the points of the "right" simplex.

Moreover, one has a partial order on vertices of such a graph $a<b$ :
\begin{equation}\label{l<}
(0***\ldots*) \,<\, ({}*1**\ldots*) \,<\, (**0*\ldots*) \,<\, \ldots
\end{equation}
on the left part, and 
\begin{equation}\label{r>}
(1***\ldots*) \,>\, ({}*0**\ldots*) \,>\, (**1*\ldots*) \,>\, \ldots
\end{equation}
on the right.
\end{Th}

\begin{Def}\label{dfn:n-simp}
The set theoretic $n$-simplex equation on the set $X$ is the following equality for the composition of $R$-maps (or correspondences) acting from right to left
\bea
\cdots \circ R_{(**0*\ldots*)} \circ R_{({}*1**\ldots*)} \circ R_{(0***\ldots*)} =
R_{(1***\ldots*)} \circ R_{({}*0**\ldots*)} \circ R_{(**1*\ldots*)} \circ \cdots \,.\nn
\eea
\end{Def}

\begin{Def}
Let us denote by $\mathfrak{S}_n(X)$ the variety of solutions for the set theoretical $n$-simplex equation in correspondences with the underlying set $X$. 
\end{Def}

Let us remind a statement from \cite{KST}.
\begin{Def}
We call an $n$-face of the $N$-cube absolutely incoming if it is not outgoing for any $n+1$-subface.
\end{Def}
\begin{Th}
\label{col}
The coloring of $n$-faces of the $N$-cube is uniquely defined by the coloring of absolutely incoming $n$-faces.
\end{Th}

Let us depict here the main arguments for the coloring theorem \ref{col}. Let us consider the coloring problem for the $(n-1)$-faces of the $N$-cube. We start with the space $C_{n+1}^{n-1}.$ Its graph $G_{n+1,n}$ has $n$-faces of the $(n+1)$-cube as vertices and as edges - such $(n-1)$-faces which are incoming for one vertex and outgoing for another. As was mentioned below the graph has two components, one of them is $I$:
\bea
(0***\ldots*) \to ({}*1**\ldots*) \to (**0*\ldots*) \to \ldots\nn
\eea
\begin{Lem}
Each absolutely incoming faces of the $(n+1)$-cube is incoming for one of $n$-faces of the simplex $I.$
\end{Lem}
Indeed, according to the definition, the absolutely incoming faces have the form $(*\ldots*\tau_i*\ldots *\tau_{j-1}*\ldots*),$ where the symbols $\tau_k$ are placed in positions with $i<j.$ Let us recall that the $\tau_k$ are defined by a sequence
\bea
\tau=(0,1,0,1,\ldots).\nn
\eea
Then it is clear that a face is incoming for the $n$-face $(*\ldots*\tau_i*\ldots*)$ from $I.$ let us remark that the same $(n-1)$-face is incoming for the $n$-face $(*\ldots *\tau_{j-1}*\ldots*)$ from the second simplex $II.$

\begin{figure}[h!]
\label{pic-I}
\center
\includegraphics[width=120mm]{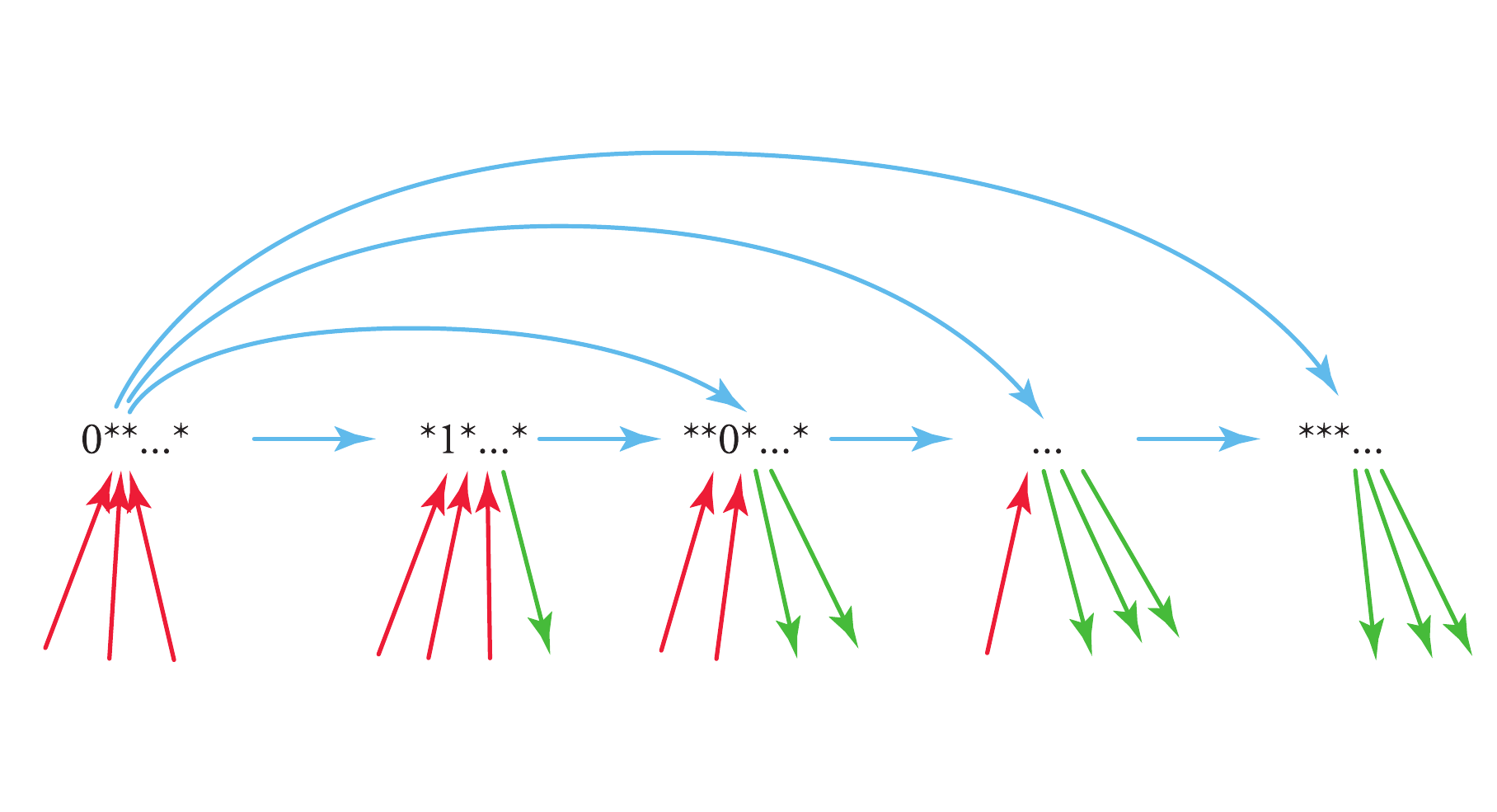}
\caption{I-configuration}
\end{figure}

We realized this structure more precisely on the picture \ref{pic-I}.

This presentation demonstrates that the $k$-th term of the complex has
\begin{itemize}
\item $n-k+1$ absolutely incoming $(n-1)$-faces, 
\item $k-1$ inner incoming $(n-1)$-faces,
\item $k-1$ absolutely outgoing $(n-1)$-faces,
\item $n-k+1$ inner outgoing $(n-1)$-faces.
\end{itemize}
This consideration connotes:
\begin{Lem}
Moving on $I$ in positive direction one can color all absolutely outgoing $(n-1)$-faces starting from the absolutely incoming ones.
\end{Lem}
\begin{Lem}
One has not other conditions then the $n$-simplex relation for coloring $(n-1)$-faces of the $(n+1)$-cube.
\end{Lem}
This follows directly from the observation that there is no common inner faces in two ways of coloring absolutely outgoing faces from the absolutely incoming ones.

Let us now demonstrate that in the $(n-1)$-faces coloring problem for higher dimension cube we do not obtain additional relations. This is demonstrated by induction, we consider another graph $\Gamma_{N,n}$ whose vertices are $(n-1)$-faces and $n$-faces in $N$-cube. They are connected by oriented edges from one to another if 
\begin{itemize}
\item $(n-1)$-face is incoming for the $n$-face;
\item for $n$-face the $(n-1)$-face is outgoing.
\end{itemize}

In fact this graph has no cycles. This implies a partial order on $(n-1)$-faces. Moreover, it is clear that starting from an absolutely incoming face one could achieve an absolutely outgoing one. 
Let us demonstrate that going throw different ways does not affect different colorings. The induction base consists in the strictly defined coloring of the absolutely incomming faces. Then let $f$ is the first $(n-1)$-face with respect to the introduced order which has different colorings throw different ways. Let $c_1$ and $c_2$ are two $n$-cube inducing different colorings of $f.$ This means that $f$ is an outgoing face to both cubes those incomming faces are colored in one way. Two $n$-cubes having a common $(n-1)$-face could be imbedded into a common $(n+1)$-cube, on which the $n$-simplex condition guaranties the coherence of outgoing faces colors. This produces a contradiction.

\subsection{Recursion}
\label{sec-rec}
Let us return to the recursion construction. Our goal is to construct a map 
\bea 
\mathfrak{r}: \mathfrak{S}_n(X)\to \mathfrak{S}_{n+1}(X^{2n}).\nn
\eea

With a solution for the $n$-simplex equation $R\in \mathfrak{S}_n(X)$ one could mount a map of the space of colorings:
\bea
\rho_n: C_N^{n-1}(X,R)\to C_N^n(X^{2n})\nn
\eea
such that a coloring for the $n$-face with an $2n$-tuple of colors is constituted by the colors of the $2n$ $(n-1)$-faces. We order the $(n-1)$-faces of an $n$-face in the following way
\bea
(i_1,\ldots,i_n,o_1,\ldots,o_n)\nn
\eea 
where $i_k$ are the incomming faces in a lexicographic order, the same numbering is supposed for the outgoing $(n-1)$-faces.
\begin{Th}
\label{th-rec}
Let $R\in \mathfrak{S}_n(X)$ then the restriction of $\rho_n$ to $C_{n+1}^{n-1}(X,R)\to C_{n+1}^n(X^{2n})$ provides a solution for the $(n+1)$-simplex equation, namely 
\bea
W=Im(\rho_n)\subset C_{n+1}^n(X^{2n})=\left(X^{2n}\right)^{n+1}\times\left(X^{2n}\right)^{n+1}\nn
\eea
is the very correspondence $W\in \mathfrak{S}_{n+1}(X^{2n}).$
\end{Th}
To demonstrate this we emphasize that the image of the restriction $\rho_n: C_{n+2}^{n-1}(X,R) \to C_{n+2}^n (X^{2n})$ 
coincides with $C_{n+2}^n (X^{2n},W)$. The $(n+1)$-simplex equation follows from the fact that the coloring of absolutely outgoing $n$-faces obtained from the coloring of absolutely incomming $n$-faces does not depend on the coloring way. The absence of higher syzygies follows from the same fact in lower dimension.

\subsection{Weight matrix $W_0$}
\label{W0}
\begin{figure}[h!]
\label{pic-YB}
\center
\includegraphics[width=60mm]{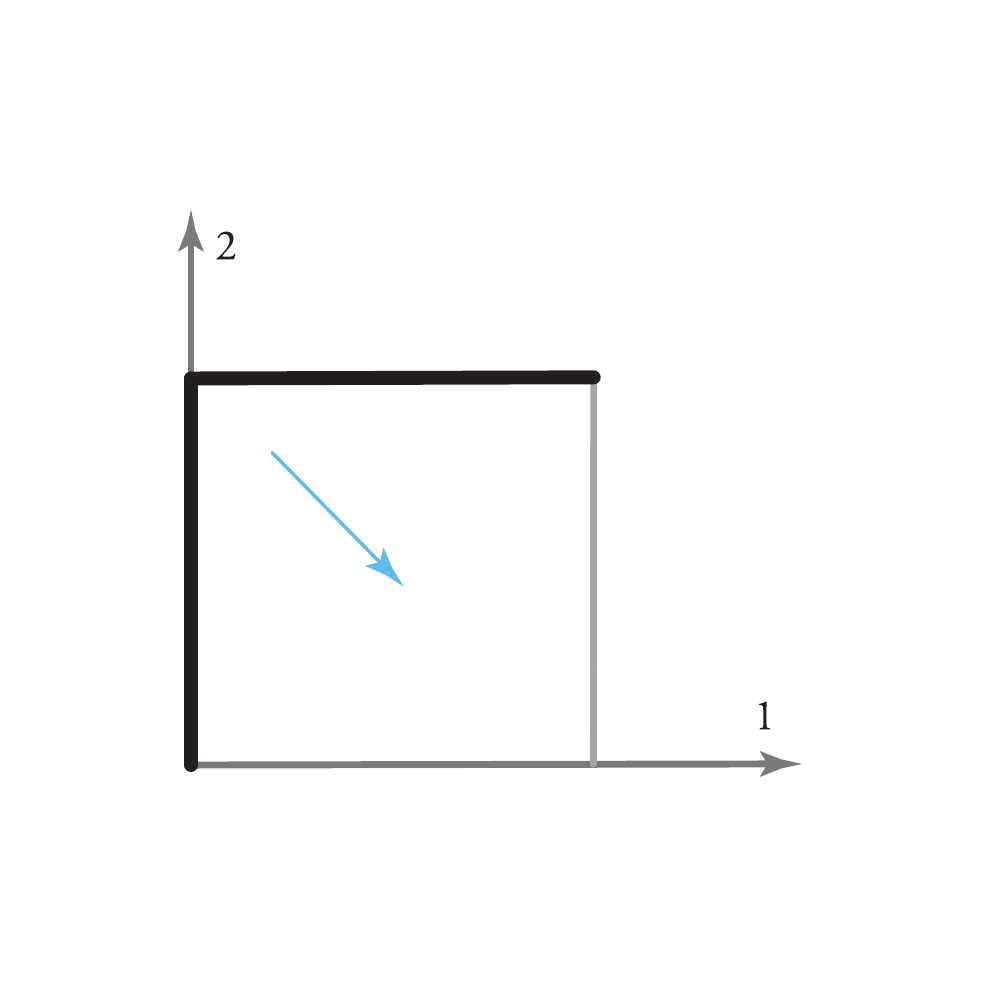}
\caption{Yang-Baxter correspondence}
\end{figure}

One could arrange the correspondences from section \ref{sec-rec} as matrix solutions for the $n$-simplex equation. For example in the case of the Ising model we are interested in the following solution for the set-theoretical Yang-Baxter equation $R$. It is represented as a map from colors of incomming edges (marked dark) to the outgoing ones on figure \ref{pic-YB}.
Let us denote the values of spins on the edges by $\{+,-\}=\{1,-1\}.$ Then the correspondence can be described as
\bea
(+,+)&\to& (+,+) \cup (-,-)\nn\\
(+,-)&\to& (+,-) \cup (-,+)\nn\\
(-,+)&\to& (+,-) \cup (-,+)\nn\\
(-,-)&\to& (+,+) \cup (-,-)\nn
\eea
Let us construct the associated matrix $R_M.$ This is defined on the space $V_X\otimes V_X$ where $V_X=\C<X>$ is the space generated by the set $X$. In our case this is the $\C^2$ space with the basis $e_1, e_2$ related to the spins $+1, -1.$ The matrix has matrix element $1$ iff the corresponding elements of $X$ belong to the correspondence. $R_M$ takes the form
\bea
R_M=\left(\begin{array}{cccc}
1 & 0 & 0 & 1\\
0 & 1 & 1 & 0\\
0 & 1 & 1 & 0\\
1 & 0 & 0 & 1
\end{array}\right).
\eea
This matrix satisfies the Yang-Baxter equation. Let us demonstrate a little bit more general statement, let $R(t)=1\otimes 1 + t \sigma \otimes \sigma$ where
\bea
\sigma=\left(\begin{array}{cc}
0 & 1\\
1 & 0
\end{array}\right)\nn
\eea
Then obviously
\bea
R_{12}(t)R_{13}(t)R_{23}(t)&=&(1+ t \sigma\otimes \sigma\otimes 1)(1+t \sigma\otimes 1 \otimes \sigma)(1+ t 1\otimes \sigma \otimes \sigma)\nn\\
&=&1+t^3 +(t+t^2)(\sigma\otimes \sigma\otimes 1+\sigma\otimes 1 \otimes \sigma+ 1\otimes \sigma \otimes \sigma).\nn 
\eea
The inverse order product gives the same symmetric expression. Moreover, we see that each composition of correspondences realizes twice. The fact that the matrix $R(1)$ has positive values guaranties that the YB equation fulfills for the correspondences.

The theorem \ref{th-rec} now furnishes a solution for the Zamolodchikov tetrahedron equation $\Phi$ for the set $X^4$ which is the set of colorings of the edges of a $2$-face of a $3$-cube. Let us introduce the space $V_f=V^{\otimes 4}$ which is the same as $\C<X^4>.$ To introduce the basis in $V_f$ we use the lexicographic order on edges of a $2$-facer: first incomming then outgoing edges. With the help of this ordering we make an identification $V_f\simeq \C^{16}$ and write an associated matrix $\Phi_M.$ The matrix elements are 1 iff the basis vectors are equated with the elements of $X^8$ lying in the correspondence.

\begin{Lem}
The associated matrix $\Phi_M$ satisfies the matrix tetrahedral equation.
\end{Lem}
To prove this one should realize the affinity between a correspondence and its matrix. Let $C_1$ and $C_2$ be two correspondences and $C_1\circ C_2$ their composition. The matrix $(C_1\circ C_2)_M$ has 1 at all positions which are in the correspondence. The matrix $(C_1)_M (C_2)_M$ differs by the values of nontrivial elements - each is the number of ways how to realize the composition of correspondences. In fact in our case this is always 2. This is done by the choice of the ``inner'' vertex in the 4-cube. This is illustrated on the picture.

\begin{figure}[h!]
\center
\includegraphics[width=70mm]{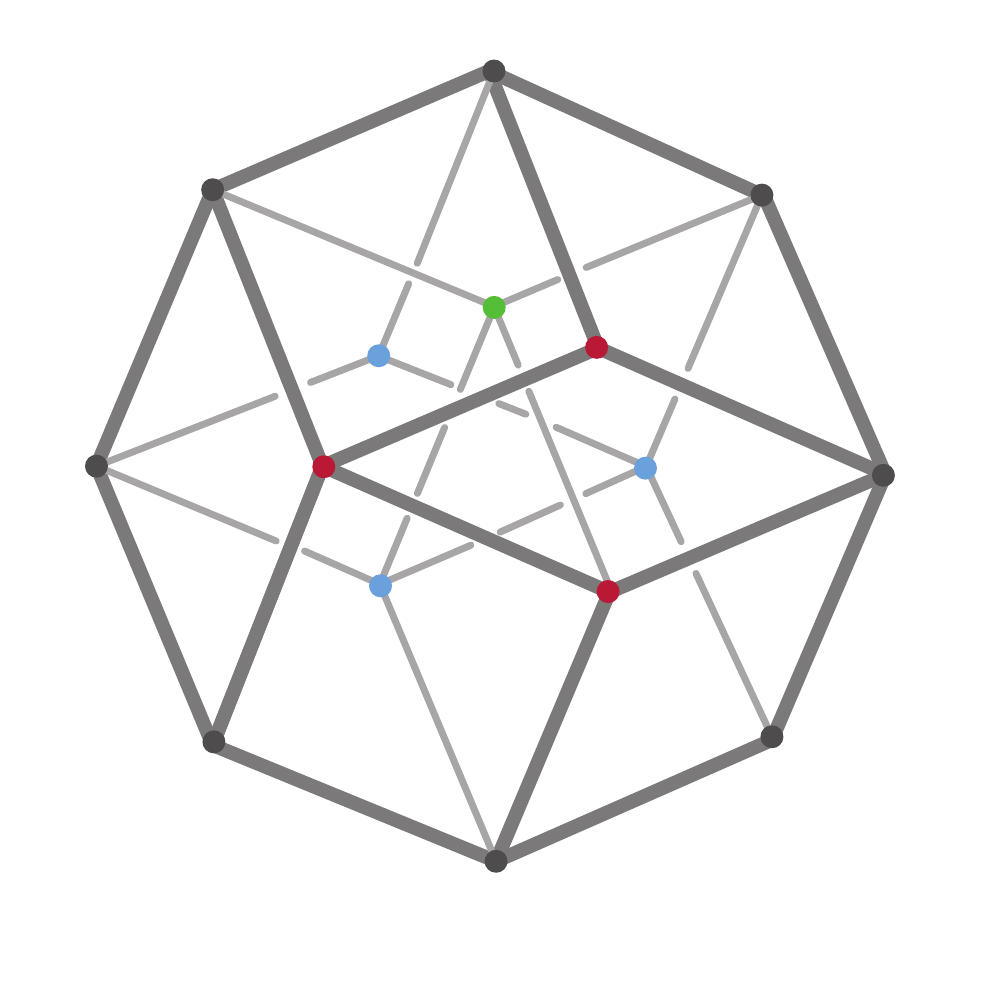}
\end{figure}

Here we represented one side of the tetrahedron equation as a process of coloring 2-faces using the relation on each 3-cube. We start with 6 2-faces in front of this projection and push behind the 3-cubes to obtain the colors of the farthest 2-faces. In our situation, the coloring is defined in fact by the colors of vertices. In addition, we see that the colors of farthest 2-faces are defined by the colors of farthest blue points. There are two ways to reach the same colors; they are determined by the color of the inner green point. 

The same can be demonstrated for the other side of the tetrahedron equation. Hence, the same set of admissible configurations of colors in both compositions of correspondences are achieved twice and hence their matrices coincide.

\section{3D Ising model transfer matrix}

\subsection{$4$-cube combinatorics}
Let us explore a $4$-cube in more details. One has the following scheme of two 3-cube sequences, denoted as L(eft) and R(ight)
\bea
&&(0***) \to (*1**) \to (**0*) \to (***1),\nn\\
&&(***0) \to (**1*) \to (*0**) \to (1***).\nn
\eea
Let us remark that there is 24 edges of the $4$-cube between 32 belonging to both sequences L and R. This could be easily seen from the fact that the L-sequence does not contain the vertex $(1010)$ and the R-sequence - $(0101)$ and hence all adjacent edges. They could be submitted in the following tables being attributed to the 3-cubes of L and R configurations respectively:
\begin{center}
\begin{tabular}{|c|c|c|c|}
\hline
0***& *1** & **0* & ***1\\
\hline
001*& 111* & 100* &10*1 \\
000* & 011*& 110*& 00*1\\
00*0 & 11*0 & 1*00& 1*11\\
01*0 & 11*1 & 0*00 & 1*01\\
0*10 & *110 & *000 & *011\\
0*11 & *100 & *001 & *111\\
\hline
\end{tabular}
\end{center}

\begin{center}
\begin{tabular}{|c|c|c|c|}
\hline
***0& **1* & *0**& 1***\\
\hline
*100& *111 & *001 & 1*01\\
*110 & *011 & *000 & 1*11\\
0*00& 0*11& 00*1 & 11*1\\
1*00& 0*10 & 10*1 & 11*0\\
01*0 & 011* & 000* & 110*\\
00*0 & 111* & 001* &100*\\
\hline
\end{tabular}
\end{center}
This choice of edges in each 3-cube correspond to a quasi-star drown on the picture for the $(***0)$- cube:
\begin{figure}[h!]
\center
\includegraphics[width=60mm]{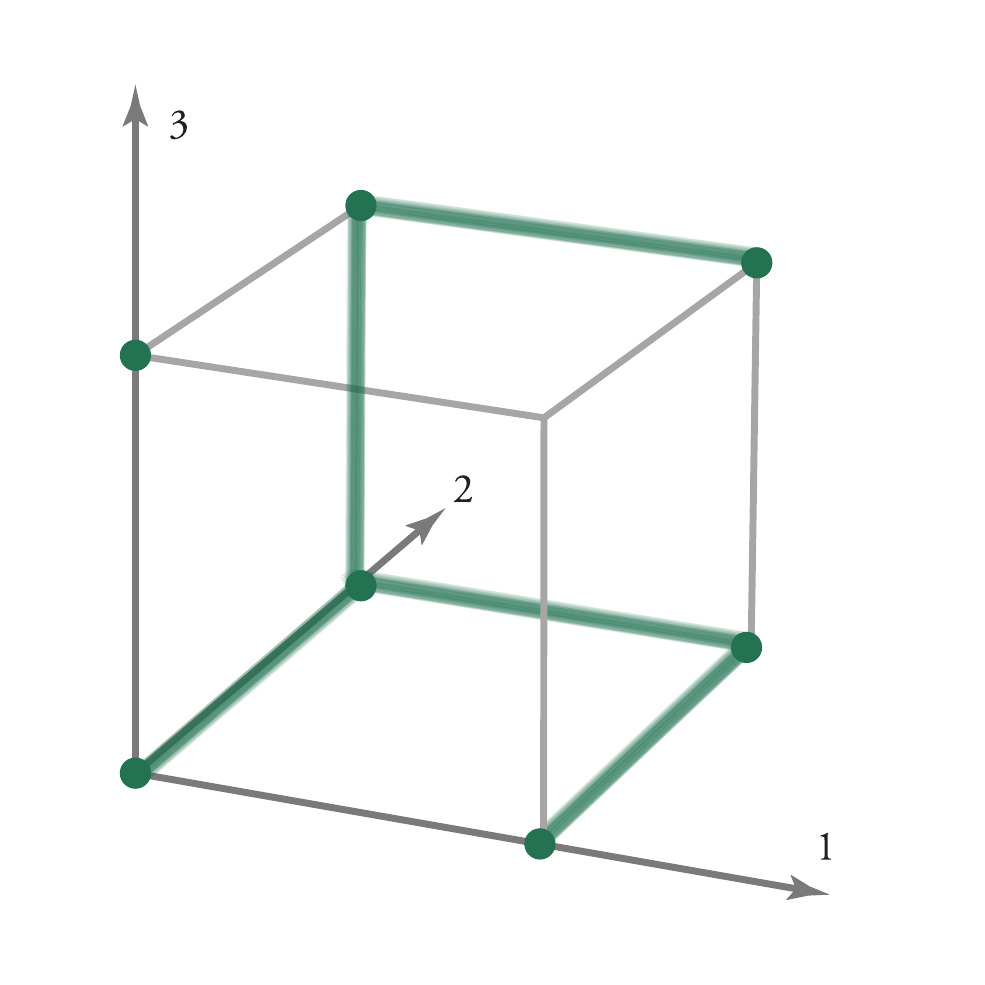}
\end{figure}

Next we observe that there is a particular choice of three edges for each 3-cube with properties:
\begin{itemize}
\item The choice of edges in a cube is defined only by the cube 3-direction, meant that it does not depend on the choice of the fixed element in a 4-cube. For example it is the same for cubes $(***0)$ and $(***1).$
\item One chooses always three edges of different directions on a 3-cube.
\item The full set of edges in configuration L given by this choice differs from the set of edges in the configuration R by the complete reversing of indices $1\leftrightarrow 4, 2\leftrightarrow 3.$
\item
Both sets have trivial intersection and their sum constitutes the whole set of edges belonging to both configurations (L and R) of 3-cubes.
\item The set of three edges of each 3-cube are what is called minimal dominant set which means that all edges of the 3-cube are adjacent to the one edge from this set.
\end{itemize}
This choice is visualized in the tables \ref{tab-LTE} and \ref{tab-RTE}
\begin{table}[h!]
\caption{Left side of the TE}
\label{tab-LTE}
\begin{center}
\begin{tabular}{|c|c|c|c|}
\hline
0***& *1** & **0* & ***1\\
\hline
01*0 & 011* & 0*00 & 00*1 \\
000* & *100 & 110* & *111 \\
0*11 & 11*1 & *001 & 1*01 \\
\hline
\end{tabular}
\end{center}
\end{table}

\begin{figure}[h!]
\center
\includegraphics[width=80mm]{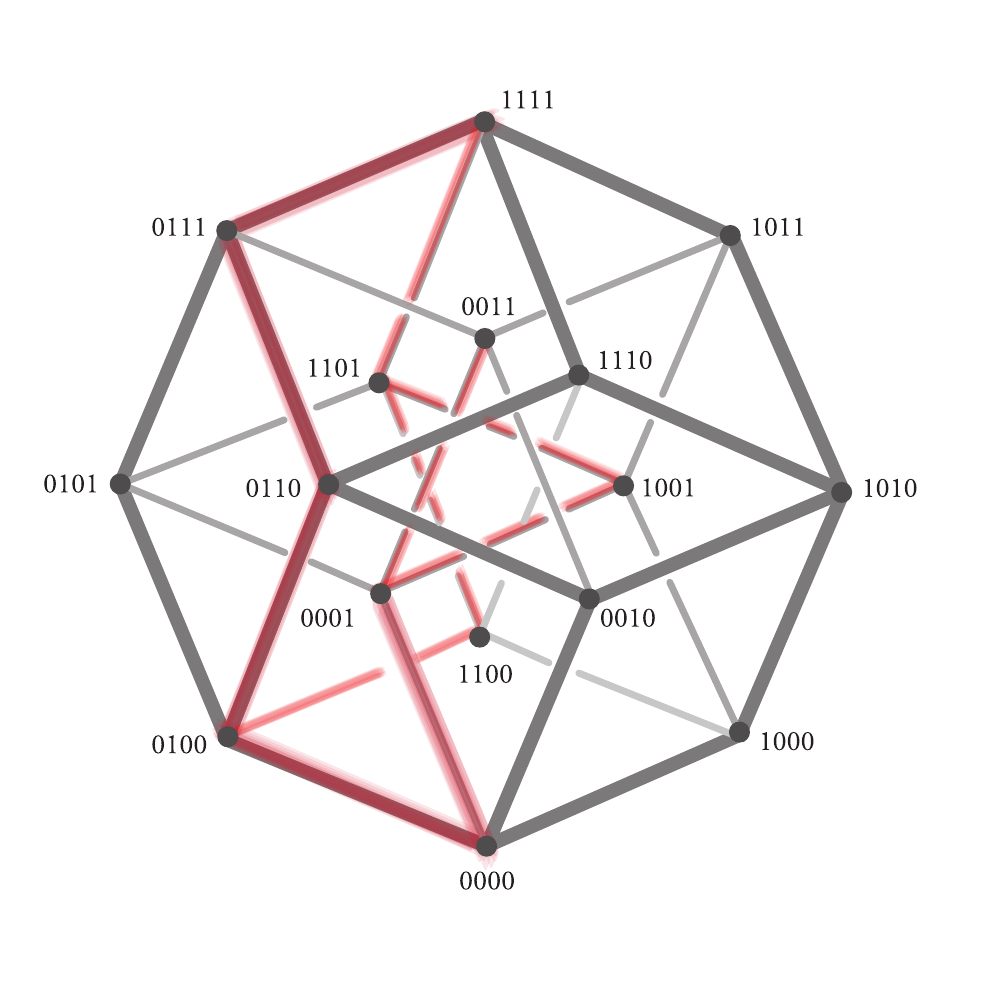}
\caption{Chosen subgraph in the 4-cube}
\label{pic-edges}
\end{figure}

\begin{figure}[h!]
\center
\includegraphics[width=80mm]{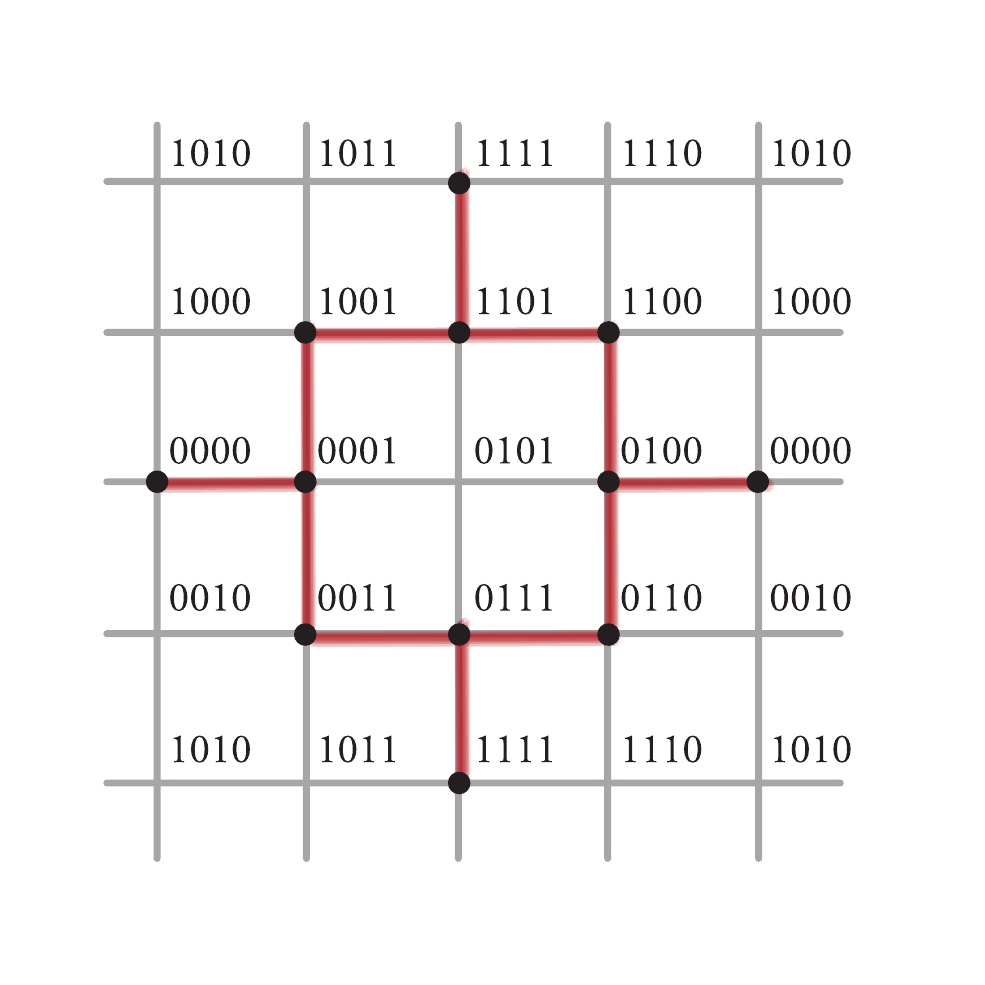}
\caption{Torus embedding}
\label{pic-torus}
\end{figure}

\begin{table}[h!]
\caption{Right side of the TE}
\label{tab-RTE}
\begin{center}
\begin{tabular}{|c|c|c|c|}
\hline
***0 & **1* & *0** & 1***\\
\hline
1*00 & 0*10 & 001* & 100* \\
00*0 & 111* & *000 & 1*11 \\
*110 & *011 & 10*1 & 11*0 \\
\hline
\end{tabular}
\end{center}
\end{table}
The picture \ref{pic-edges} illustrates the table \ref{tab-LTE}.
We would like to propose another interpretation for this subgraph. In fact the 4-cube is known to be the graph of genus 1, it could be embedded into a 2-torus, such that its vertices match the vertices of the periodic 4-order 2-dimensional lattice in the torus. In this notation the subgraph of chosen 12 edges is represented by the picture \ref{pic-torus}
This subgraph $\Gamma'$ possesses the property with respect to the graph $\Gamma$ of the 4-cube appreciated in coding theory, if $d_{\Gamma'}(V_1,V_2)\le 2$ then $d_{\Gamma}(V_1,V_2)\le 2$ where $d_{\Gamma}(V_1,V_2)\le 2$ is the distance in the graph $\Gamma$ between points $V_1$ and $V_2.$

\subsection{Twisted tetrahedron equation}
With the help of the auxiliary weight matrix $W_0$ from section \ref{W0} and observation of the preceding section we now compose a weight matrix for the 3D Ising model being guided by the following idea: in place of each nontrivial element of $W_0$ we put a weight $exp(t(\sigma_1+\sigma_2+\sigma_3))$ where $\{\sigma_i\}$ are the spin values of the chosen edges of the 3-cube. We denote the new matrix by $W(i,j,k).$
\begin{Rem}
The fact that we always choose three edges of different directions provides $W$ be the weight matrix for the 3D Ising model. 
\end{Rem}

Now let us analyze the way how we choose the edges according to the tables \ref{tab-LTE} and \ref{tab-RTE}. This rule may well take a more analytical form. Let us introduce coordinates on our combinatorial objects in table \ref{tab-ind}.
\begin{table}[h!]
\caption{Indexing}
\label{tab-ind}
\begin{center}
\begin{tabular}{p{2cm} | p{12cm}}
3-cubes & numbers of $*$-positions $i,j+1,k+2.$\\
\hline
2-faces & numbers of $*$-positions inside a 3-cube $m, n+1,$ and the direction $d$ ($0$ - for incomming and $1$ for outgoing faces).\\
\hline
edges & $l$ - number of $*$-position in a 2-face, $s$ - direction.
\end{tabular}
\end{center}
\end{table}
\begin{Lem}
\label{func}
In the notation of the table \ref{tab-ind} the choice of 3 egdes of a 3-cube is given by the formula
\bea
l&=&m (i+j+1)+n(j+k+1)+(i+j+d),\nn\\
s&=&m((d+1)(i+j)+1)+n(d(j+k)+1)+(d(i+j)+(j+k)+1).\nn
\eea
Here we get an edge from each 2-face of a 3-cube.
\end{Lem}
The proof is straightforward.

\begin{Def}
\label{weight}
The weight function $W(i,j+1,k+2)$ is defined as a matrix with matrix elements indexed by 6 2-faces of a 3-cube
\bea
&&W(i,j+1,k+2)_{(m_1,n_1,d_1),(m_2,n_2,d_2),(m_3,n_3,d_3)} ^{(m_4,n_4,d_4),(m_5,n_5,d_5),(m_6,n_6,d_6)}\nn\\
&&=W_0(i,j+1,k+2)_{(m_1,n_1,d_1),(m_2,n_2,d_2),(m_3,n_3,d_3)} ^{(m_4,n_4,d_4),(m_5,n_5,d_5),(m_6,n_6,d_6)} \nn\\
&&\times \exp(t\sum_{u=1}^6 \sigma_{l(m_u,n_u,d_u,i,j,k),s(m_u,n_u,d_u,i,j,k)})
\eea
with functions $l,s$ given by the lemma \ref{func}.
\end{Def}
\begin{Rem}
The definition means that for only admitted combinations of states of edges of each 2-face of the 3-cube (this is given by the associated matrix $W_0$ for the correspondence) we choose an edge for each 2-face and calculate the exponential of their spins sum.
\end{Rem}
\begin{Lem}
The matrix $W(i,j+1,k+2)$ gives a weight matrix for the 3D Ising model for each choice of parameters $i,j$ and $k.$ This means that the partition function $Z(t)$ can be obtained as a product
\bea
Z(t)=\prod_{(\alpha,\beta,\gamma)\in\Lambda^*} W_{\alpha\beta\gamma}(i,j,k)
\eea
over the dual lattice $\Lambda^*.$
\end{Lem}

\begin{Rem}
Let us consider the symmetry transformation $T$ for the 4-cube consisting in the indices exchange $1 \leftrightarrow 4, 2\leftrightarrow 3.$ After such transposition the 3-cubes are changed, for example one has a mutation $(***1)\leftrightarrow (1***)$. The matrix elements of matrices $W_{\alpha, \beta, \gamma}(i,j,k)$ are changed as follows: the incoming 2-faces stay incoming but with altered order, the first incoming 2-face exchanges with the third. The edges inside a 2-face change completely $l\to 1-l$ and $s\to 1-s.$ Let us denote this linear transformation in $V_f$ as an operator $A.$
The transformation $T$ hence expresses as follows for the particular example
\bea
T(W_{356}(2,3,4))=A_1 A_2 A_3 W_{321}(2,3,4) A_1 A_2 A_3=W^A_{321}(2,3,4) \nn
\eea
\end{Rem}
\begin{Th}
The weight matrix from definition \ref{weight} satisfies an equation
\bea
&&W_{653}^A(1,2,3) W_{642}^A(1,2,4) W_{541}^A(1,3,4) W_{321}^A(2,3,4)\nn\\
&&=
W_{356}(2,3,4) W_{246}(1,3,4) W_{145}(1,2,4) W_{123}(1,2,3).
\label{TTE}
\eea
\end{Th}
\begin{Rem}
We call the equation \ref{TTE} the {\bf twisted tetrahedron equation} for the reason that is could be interpreted in a very similar form:
\bea
&&P_{16} P_{25} W_{123}^A(1,2,3) W_{145}^A(1,2,4) W_{246}^A(1,3,4) W_{356}^A(2,3,4)P_{16} P_{25} \nn\\
&&=
W_{356}(2,3,4) W_{246}(1,3,4) W_{145}(1,2,4) W_{123}(1,2,3).
\label{TTE2}
\eea
where $P_{ij}$ are just transpositions in the pair of $i$-th and $j$-th spaces.
\end{Rem}
The proof is quite straightforward; the principal thing is the property of our choice of edges on 2-faces. The transformation $T$ converts one choice to the other. Then one should realize how this transformation acts on the matrix multiplication. In fact, it does not affect the order of multipliers but changes the indexing.

\section{Gray code and related problems}
In fact the hypercube combinatorics is strongly related to the coding theory, for example the hamiltonian cycles on $n$-cube, which are the cycles without self-intersections passing throw all the vertices of the graph, are nothing but the realizations of the Gray code \cite{GC}, which is the method to encode the states of some system such that two successive values differ in only one bit. This code is widely applied for minimizing errors while transforming analogous signals to digital, in television, in computer mice etc.

In our case the way to choose the edges in each 3-cube provides a graph in a 4-cube which is combinatorically a 3-simlex with additional points on its edges. For example it could be interpreted as a symmetric redundancy code with distance $2$ on a 4-cube. It is represented by the following set of admitted code combinations $\{(1111), (0011), (1001), (0000), (1100), (0110)\}$ which are identified with the points on the tetrahedron edges. In fact, this technique produces some code for hypercubes which can be computed locally. This is of interest from the computational complexity point of view. 

Another application of this construction is influenced by the edge-edge-dominance property of the constructed subgraph of the hypercube. 

The Gray code with correction is related to the so-called coil-in-the-box problem or more generally $(n,k)$-chain codes. In fact the exploited subgraph in the 4-cube contains an induced cycle 
\bea
\begin{array}{cccc}
(0100) & (1100) & (1101) & (1001) \\
(0001) & (0011) & (0111) & (0110)
\end{array}\nn
\eea
which provides a $(4,2)$-code.
\section{Conclusion}
The result of this work can be seen as an intermediate step in applications of algebraic integrable structures for the solution of the 3D Ising model. However, we constructed a vertex-model description for the 3D Ising model and demonstrated its weight matrix satisfies the twisted version of the tetrahedron equation. Moreover, the combinatorial substance of this solution provides its own interest for the fundamental problems in coding theory.

\end{document}